\documentclass[reprint,superscriptaddress,amsmath,aps,showpacs,pra,nofootinbib,longbibliography]{revtex4-2}
\usepackage{amssymb}
\usepackage{appendix}
\usepackage{verbatim}
\usepackage{float}
\usepackage{color}
\usepackage{textcomp}
\usepackage{gensymb}
\usepackage{dsfont}
\usepackage{hyperref}
\hypersetup{
     colorlinks   = true,
     citecolor    = blue
}
\usepackage{siunitx}
\usepackage{graphicx}
\usepackage{dcolumn}
\usepackage{bm}
\usepackage{braket}
\usepackage{empheq}
\usepackage{nicefrac} 
\usepackage[dvipsnames]{xcolor}

\RequirePackage{color}
\usepackage[normalem]{ulem}

\begin{document}

\title{Tunable topological narrow bands in twisted bilayer-trilayer graphene}
\author{Dong Wang}
\affiliation{Key Laboratory of Artificial Micro- and Nano-structures of Ministry of Education and School of Physics and Technology, Wuhan University, Wuhan 430072, China}
\author{Federico Escudero}
\affiliation{IMDEA Nanoscience, Faraday 9, 28049 Madrid, Spain}
\author{Zhen Zhan}
\email{zhenzhanh@gmail.com}
\affiliation{Key Laboratory of Artificial Micro- and Nano-structures of Ministry of Education and School of Physics and Technology, Wuhan University, Wuhan 430072, China}
\affiliation{IMDEA Nanoscience, Faraday 9, 28049 Madrid, Spain}
\author{Shengjun Yuan}
\email{s.yuan@whu.edu.cn}
\affiliation{Key Laboratory of Artificial Micro- and Nano-structures of Ministry of Education 
and School of Physics and Technology, Wuhan University, Wuhan 430072, China}
\affiliation{School of Artificial Intelligence, Wuhan University, Wuhan 430072, China}
\affiliation{Wuhan Institute of Quantum Technology, Wuhan, 430206, China}

\date{\today}

\begin{abstract}
We investigate the low-energy band structure and topology of twisted bilayer--trilayer graphene with four stacking configurations: AB--ABC, BA--ABC, AB--ABA, and BA--ABA. Using both tight-binding and continuum models, we first establish that the two approaches show good agreement in the band structure in low-energy regime. We then study the evolution of the flat bands and their valley Chern numbers as functions of twist angle, perpendicular electric field, and the self-consistent Hartree potential. At relatively large twist angles and under electric field, we find a topological transition between the narrow bands, with the total Chern number of the flat bands following the Chern number sum rules derived from the chiral-limit description. We also observe another type of topological transition when the flat bands hybridize with adjacent remote bands, where gap closing and reopening processes lead to Chern number and charge density transfer. By constructing topological phase diagrams in the space of twist angle and electric field, we show that the perpendicular electric field provides an efficient tuning knob for controlling the stability and transitions of the Chern bands. Finally, we find that the Hartree potential mainly induce weak band shifts and reshaping in the narrow bands. However, with a combination of Hartree potential and the electric fields, the narrow bands show rich topological phase diagram. Our results clarify the interplay between the stacking, twist angle and electric field in manipulating the narrow bands and their topology in twisted bilayer--trilayer graphene, and provide guidance for engineering topological narrow bands with tunable Chern numbers in realistic twisted multilayer graphene systems.
\end{abstract}

\maketitle
\section{Introduction}

Twisted graphene systems have attracted considerable attention 
as a platform for exploring novel correlated and topological phases of matter, owing to their relatively simple structures and multiple tunable degrees of freedom \cite{andrei2021marvels,yang2026twisting,ju2024fractional}.
Since the breakthrough discovery of correlated insulating states and unconventional superconductivity in twisted bilayer graphene \cite{caoCorrelatedInsulatorBehaviour2018,caoUnconventionalSuperconductivityMagicangle2018}, a variety of novel states and corresponding phenomena have been proposed or measured in moiré graphene systems, including topological bands \cite{songAllMagicAngles2019a,liuPseudoLandauLevel2019a,ahnFailureNielsenNinomiyaTheorem2019,poFaithfulTightbindingModels2019,rademakerTopologicalFlatBands2020}, Chern insulator states (CIs) \cite{nuckollsStronglyCorrelatedChern2020,wang2026programmable}, quantum anomalous Hall effect (QAHE) \cite{serlinIntrinsicQuantizedAnomalous2020}, and fractional Chern insulator states (FCIs)~\cite{abouelkomsanParticleHoleDualityEmergent2020,repellinChernBandsTwisted2020,xieFractionalChernInsulators2021a,dongManybodyGroundStates2023a}.
Crucially, the electronic structure and topology of these systems can be efficiently controlled by several experimentally accessible tuning knobs, including the twist angle \cite{bistritzerMoireBandsTwisted2011,lopesdossantosGrapheneBilayerTwist2007,kimTunableMoireBands2017}, external electric field \cite{parkGatetunableTopologicalFlat2020a}, Coulomb interaction \cite{luSuperconductorsOrbitalMagnets2019}, strain~\cite{escudero2026straintronics}, stacking configuration \cite{koshinoBandStructureTopological2019,parkGatetunableTopologicalFlat2020a}, and layer number \cite{phong2025coulombinteractionstabilizedisolatednarrow,waters2024topological,zhang2025layer,chen2026layer}.

The FCIs have been experimentally observed in two classes of moiré materials: twisted MoTe$_2$ bilayer \cite{cai2023signatures,zeng2023thermodynamic,xu2023observation,redekop2024direct} and rhombohedral-stacked multilayer graphene \cite{lu2024fractional,xie2025tunable,lu2025extended}. These two systems fractionalize from doping narrow bands with valley Chern number $|C|=1$. Recently, twisted multilayer graphene systems have been proposed to be promising platforms for realizing FCIs in flat bands with higher Chern numbers $|C|>1$~\cite{liuGateTunableFractionalChern2021,yangFlatBandsHigh2023,phong2025coulombinteractionstabilizedisolatednarrow,dongObservationIntegerFractional2025}. 
Unlike FCIs in $|C|=1$ flat bands~\cite{neupertFractionalQuantumHall2011, sunNearlyFlatbandsNontrivial2011,regnaultFractionalChernInsulator2011,shengFractionalQuantumHall2011,tangHighTemperatureFractionalQuantum2011}, a single isolated band with a higher Chern number, in general, cannot be simply mapped onto multiple decoupled Landau levels, but rather, is analogous to higher Landau levels ~\cite{liuFractionalChernInsulators2012,wangFractionalQuantumHall2012,trescherFlatBandsHigher2012,yangTopologicalFlatBand2012,wuBlochModelWave2013,sterdyniakSeriesAbelianNonAbelian2013,mollerFractionalChernInsulators2015}. 
This distinction gives rise to richer states in higher-Chern number bands, for example, stabilize fractional states at generalized Jain bosonic and fermionic fillings \cite{ju2024fractional,cao2025fractional}.

The origin of higher Chern numbers in twisted chirally stacked multilayer graphene has been discussed in detail within continuum and chiral-model descriptions~\cite{liuQuantumValleyHall2019,ledwithFamilyIdealChern2022,wangHierarchyIdealFlatbands2022}. 
In these models, the enhanced Chern number arises from a hierarchical (or wave-function-exchange) mechanism in which the additional outer graphene layers modify the low-energy moiré flat-band wave functions of the twisted subsystem. 
In the chiral limit, these flat bands can further exhibit ideal quantum geometry, which is favorable for stabilizing FCIs~\cite{ledwithFamilyIdealChern2022,wangHierarchyIdealFlatbands2022,dongManybodyGroundStates2023a}. 
As a result, for a twisted system composed of $M$- and $N$-layer graphene stacks, the total Chern number of the two low-energy flat bands, in a given spin and valley sector, is determined by the layer numbers and the relative stacking chirality, taking the form $\pm(M-N)$ for the same stacking chirality and $\pm(M+N-2)$ for the opposite stacking chirality, with the overall sign depending on the valley and convention~\cite{liuQuantumValleyHall2019,ledwithFamilyIdealChern2022}. These general results suggest that twisted multilayer graphene provides a systematic route to engineering flat bands with tunable and potentially large Chern numbers. 
However, these conclusions were mainly obtained under idealized conditions, such as the chiral limit or simplified continuum descriptions, and also lack the details of the tunability of the topological bands with tuning knobs, i.e. twist angle, external electric field, and Coulomb interaction. 

Previous studies have discussed the gate-tunable band structures and topology of twisted monolayer-bilayer graphene and twisted double bilayer graphene~\cite{koshinoBandStructureTopological2019,parkGatetunableTopologicalFlat2020a}. Among the twisted $M$--$N$ multilayer graphene system, experimental results show that twisted bilayer-trilayer graphene has the largest extent of symmetry-broken phases changing with filling and electric fields \cite{waters2024topological}, and more interestingly, the moiré-driven topological electronic crystal \cite{su2025moire}.  
Therefore, in this work, we focus on twisted bilayer-trilayer graphene and investigate how its low-energy band structure and topology are manipulated by several tuning knobs. We first establish the agreement between the tight-binding (TB) and continuum models (CM) in the low-energy window and analyze the stacking-dependent band features. 
We then calculate the valley Chern numbers of the conduction and valence flat bands as a function of twist angle and identify the gap-closing processes responsible for topological transitions. 
At relatively large twist angles, the total Chern number of the two flat bands follows the hierarchy expected from the chiral-limit theory, whereas additional transitions can occur when the flat bands hybridize with nearby remote bands. 
We further study the effect of a perpendicular electric field, which provides an efficient way to isolate or rehybridize the layer-polarized flat bands that leads to rich Chern number phase diagrams. 
Finally, we include the self-consistent Hartree potential at the mean-field level and examine its filling-dependent influence on the flat-band dispersion and topology.

Our results show that twisted bilayer--trilayer graphene hosts tunable topological narrow bands whose Chern numbers are controlled by the interplay between twist angle, stacking configuration, electric field, and Hartree potential. 
In particular, the ABC-based structures tend to exhibit stable isolated flat-band manifolds. On the contrary, the ABA-based structures are highly tunable due to an additional monolayer-like band close to the flat-band manifold. 
The perpendicular electric field provides the dominant tuning knob for changing the band topology, while the self-consistent Hartree correction mainly induces weak filling-dependent band renormalization, without generally changing the Chern numbers unless a relevant gap is closed. Moreover, due to the interplay between the twist angle and electric field, two types of topological transitions are observed: one type of transition occurs between the narrow bands, of which the total Chern number of these two narrow bands satisfies the sum rules derived from the chiral model; the other type of transition happens between the narrow band and its adjacent remote band. These findings clarify the stacking- and field-tunable topology of twisted bilayer--trilayer graphene and provide guidance for engineering higher Chern number flat bands in realistic twisted multilayer graphene systems.

The paper is organized as follows: In Sec.~\ref{sec:method}, we describe the atomic structures and stacking configurations of twisted bilayer--trilayer graphene, the TB and CM approaches, together with the numerical methods used to calculate the valley Chern number and the self-consistent Hartree potential.
In Sec.~\ref{sec:results}, we present and discuss the band structures and their topological evolution as functions of twist angle, perpendicular electric field, and carrier filling.
Finally, Sec.~\ref{sec:conclusion} summarizes our main conclusions.

\section{Numerical methods}
\label{sec:method}
In this part, we first introduce the moiré structures and then the numerical details adopted in this work. 

\subsection{Atomic structures}
\label{sec:struct}
\begin{figure}[t!]
    \centering
    \includegraphics[width=0.9\linewidth]{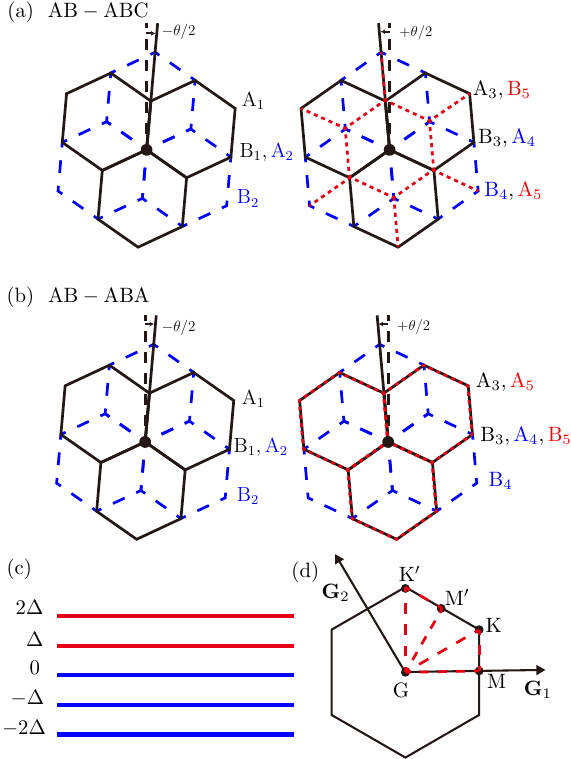}
    \caption{Schematic atomic structures and moiré geometry of twisted bilayer--trilayer graphene. (a) AB-stacked bilayer graphene and ABC-stacked trilayer graphene, rotated by $-\theta/2$ and $+\theta/2$, respectively, to form the AB--ABC configuration. (b) Corresponding construction of the AB--ABA configuration. Atoms sharing the same in-plane position are listed together, while the subscripts label the graphene layers. (c) Layer-dependent electrostatic potential used to model a perpendicular electric field, with onsite energies $(2\Delta,\Delta,0,-\Delta,-2\Delta)$ assigned from the top to the bottom layer. (d) Moiré Brillouin zone, showing the reciprocal lattice vectors $\mathbf{G}_1$ and $\mathbf{G}_2$ and the high-symmetry points $\mathrm{G}$, $M$, $M'$, $K$, and $K'$. }
    \label{fig:struct}
\end{figure}

We consider twisted bilayer--trilayer graphene composed of a Bernal-stacked bilayer placed on top of either an ABC- or ABA-stacked trilayer. The five graphene layers are labeled from top to bottom by $l=1,\ldots,5$, with layers 1 and 2 belonging to the bilayer and layers 3--5 belonging to the trilayer. Each graphene layer contains two sublattices, denoted by $A_l$ and $B_l$.
We consider both AB and BA bilayers. 
In our convention, the vertically aligned dimer sites of the AB bilayer are $B_1$ and $A_2$, whereas those of the BA bilayer are $A_1$ and $B_2$.  
Changing AB into BA reverses the internal stacking orientation of the bilayer and interchanges the dimer and nondimer sublattices.
The two trilayer configurations considered here are rhombohedral ABC stacking and Bernal ABA stacking.
The ABC trilayer is a uniformly chiral, rhombohedrally stacked multilayer. 
By contrast, the ABA trilayer contains two interfaces with opposite stacking orientations. 
Its low-energy electronic structure can instead be decomposed into monolayer-like and bilayer-like sectors~\cite{minChiralDecompositionElectronic2008,koshinoGateinducedInterlayerAsymmetry2009a}.
This distinction will be important below, since the monolayer-like sector of the ABA trilayer produces an additional Dirac band close to the moiré narrow bands.
To construct the twisted structures, the bilayer and trilayer substacks are rigidly rotated by $-\theta/2$ and $+\theta/2$, respectively, about a common threefold rotation center. 
The AB--ABC and AB--ABA structures are illustrated in Fig.~\ref{fig:struct}(a) and \ref{fig:struct}(b), respectively. 
The BA--ABC and BA--ABA configurations are obtained by reversing the internal stacking orientation of the bilayer while leaving the trilayer stacking sequence unchanged.

For the $C_{3z}$-symmetric relative registry adopted here, the full twisted structure preserves threefold rotation symmetry about the out-of-plane axis. 
Time-reversal symmetry is also preserved when both valleys are included. 
By contrast, a single-valley Hamiltonian is not invariant under ordinary time reversal. 
The bilayer and trilayer substacks contain different numbers of layers and cannot be interchanged by a crystalline symmetry. 
Consequently, the complete $2+3$ structure generally lacks a horizontal mirror symmetry and the in-plane twofold rotations that would exchange the top and bottom substacks. 
It also lacks $C_{2z}$ symmetry for the stacking registries considered here. 
The absence of these twofold symmetries permits nondegenerate bands along generic high-symmetry paths within a single valley.

We define the primitive lattice vectors of monolayer graphene as  
\begin{equation} 
\mathbf{a}_1=a(1,0),\qquad \mathbf{a}_2=a\left(1/2,\sqrt{3}/2\right), \end{equation}  
where $a=2.46~\text{\AA}$ is the graphene lattice constant. 
For the atomistic tight-binding calculations, we use commensurate moiré supercells with twist angles satisfying~\cite{lopesdossantosGrapheneBilayerTwist2007,meleCommensurationInterlayerCoherence2010} 
\begin{equation} 
\cos\theta= \frac{3m^2+3m+1/2} {3m^2+3m+1}, 
\label{eq:commensurate_angle} 
\end{equation}  
where $m$ is a positive integer. 
The corresponding moiré superlattice vectors are 
\begin{align} 
\mathbf{L}_1 &= m\mathbf{a}_1+(m+1)\mathbf{a}_2,\\\nonumber 
\mathbf{L}_2 &= -(m+1)\mathbf{a}_1+(2m+1)\mathbf{a}_2. 
\label{eq:moire_lattice_vectors} 
\end{align} 
The reciprocal lattice vectors $\mathbf{G}_1$ and $\mathbf{G}_2$ are defined by 
\begin{equation} 
\mathbf{G}_i\cdot\mathbf{L}_j=2\pi\delta_{ij}. 
\end{equation}  
The moiré Brillouin zone (mBZ) and the high-symmetry points $\mathrm{G}$, M, $\mathrm{M^\prime}$, K, and $\mathrm{K^\prime}$ are shown in Fig.~\ref{fig:struct}(d).

To model a uniform perpendicular electric field, we introduce a linear layer-dependent onsite potential. 
With the layers ordered from top to bottom, the electrostatic potentials are chosen as 
\begin{equation} 
(U_1,U_2,U_3,U_4,U_5) = (2\Delta,\Delta,0,-\Delta,-2\Delta), 
\label{eq:layer_potential} \end{equation}  
where $\Delta$ is the potential-energy difference between adjacent layers. 
The corresponding potential profile is illustrated in Fig.~\ref{fig:struct}(c).

All atomic structures used in this work are generated by rigidly rotating ideal graphene layers.
The interlayer separation between adjacent layers is fixed at $h=3.349~\text{\AA}$, and neither in-plane lattice relaxation nor out-of-plane corrugation is included explicitly in the tight-binding model. 
In the continuum model, the leading effect of lattice relaxation is incorporated phenomenologically through unequal AA- and AB-region tunneling amplitudes, $\gamma_{AA}<\gamma_{AB}$ \cite{namLatticeRelaxationEnergy2017}. We leave the comprehensive analysis of lattice relaxation to future studies.

\subsection{Tight-binding model}

We extend the tight-binding model for twisted bilayer graphene (TBG) that includes only the carbon $p_z$ orbitals to the multilayer case \cite{trambly2012numerical}.
The Hamiltonian of the graphene moiré system is given by
\begin{equation}
H=\sum_i \varepsilon_i c^{\dagger}_{i} c_{i}+ \sum_{\langle i,j \rangle} t_{ij} c^{\dagger}_{i} c_{j},
\end{equation}
where $c_i$ ($c_i^{\dagger}$) is the annihilation (creation) operator for an electron in state $i$, $\varepsilon_i$ denotes the on-site potential, $\langle i,j \rangle$ represents a summation over lattice sites with $i \neq j$, and $t_{ij}$ is the hopping integral between orbitals at sites $i$ and $j$.
To simplify the problem, we only consider the nearest-neighbor intralayer hopping and neglect the interlayer hopping between non-adjacent layers, which captures the main feature of the systems \cite{guinea2019continuum,wu2021lattice}.
The interlayer hopping is described by the Slater–Koster formalism.
According to this approach, the hopping integral $t_{ij}$ between two $p_z$ orbitals located at positions $\mathbf{r}_i$ and $\mathbf{r}_j$ is expressed as \cite{trambly2012numerical}
\begin{equation}
t_{ij}=n^2 V_{pp\sigma}(r_{ij})+(1-n^2)V_{pp\pi}(r_{ij}),
\end{equation}
where $r_{ij}=|\mathbf{r}_i-\mathbf{r}_j|$ is the distance between sites $i$ and $j$, and $n=z_{ij}/r_{ij}$ is the direction cosine of the relative displacement vector along the $z$ axis.
The Slater–Koster parameters $V_{pp\sigma}$ and $V_{pp\pi}$ are chosen as
\begin{align}
V_{pp\pi}(r_{ij}) &=-t_0 e^{q_{\pi}(1-r_{ij}/d)} F_c(r_{ij}), \\\nonumber
V_{pp\sigma}(r_{ij}) &= t_1 e^{q_{\sigma}(1-r_{ij}/h)} F_c(r_{ij}),
\end{align}
where $d=1.42~\text{\AA}$ and $h=3.349~\text{\AA}$ denote the nearest-neighbor in-plane distance and the interlayer spacing, respectively. 
The parameters $t_0$ and $t_1$ are adjustable and are typically reparameterized to fit experimental observations. In this work, we set $t_0=2.8$ eV, $t_1=0.44$ eV, which gives a magic angle of $1.05^\circ$ in the TBG~\cite{kuang2021collective}. 
The decay parameters satisfy $q_{\sigma}=7.43$ and $q_{\pi}=3.15$.
The smooth cutoff function is defined as $F_c(r)=\left[1+e^{(r-r_c)/l_c}\right]^{-1}$ with $l_c=0.265~\text{\AA}$ and $r_c=6.14~\text{\AA}$. For distances larger than 5~\text{\AA}, we assume the interlayer hopping is zero. All the TB calculations are performed in the TBPLaS simulator \cite{li2023tbplas}.

To identify the valley character of the bands in the TB model, we employ the valley operator~\cite{ramiresImpurityinducedTriplePoint2019a,colomesAntichiralEdgeStates2018,ramiresElectricallyTunableGauge2018}, defined as
\begin{equation}
    \hat{V}_z=\frac{i}{3\sqrt{3}}\sum_{\langle i,j\rangle}\eta_{ij}\sigma_z^{ij}c_i^\dagger c_j,
\end{equation}
where $i,j$ denote next-nearest-neighbor sites, $\eta_{ij}=\pm1$ represents the sign convention for clockwise and counterclockwise hopping, and $\sigma_z^{ij}$ is the Pauli matrix associated with the sublattice degree of freedom. 
The expectation value of this operator ranges from $+1$ to $-1$.
The value close to $+1$ and $-1$ corresponding to the two inequivalent valleys provides a valley polarization index.

\subsection{Continuum model}

To construct the continuum model of multilayer graphene, we adopt the effective-mass description in the Slonczewski-Weiss-McClure parameterization \cite{jungAccurateTightbindingModels2014,partoensGrapheneGraphiteElectronic2006,phong2025coulombinteractionstabilizedisolatednarrow}.
The Hamiltonian of twisted trilayer-bilayer stack at valley $\xi=\pm1$ has the following form
\begin{equation}
    \mathbb{K}_{3+2,\xi}(\mathbf{k})=\begin{pmatrix}
        \mathbb{K}_{3,\xi}(\mathbb{R}[\theta/2]\mathbf{k}) & \mathbb{T}_\xi \\
        \mathbb{T}_\xi^\dagger & \mathbb{K}_{2,\xi}(\mathbb{R}[-\theta/2]\mathbf{k})
    \end{pmatrix},
\end{equation}
where $\mathbb{K}_{3,\xi}$, $\mathbb{K}_{2,\xi}$ are the Hamiltonian of trilayer and bilayer graphene, and $\mathbb{T}_{\xi}$ is the tunneling matrix between the substacks.
We neglect the hopping between non-adjacent layers to isolate the dominant interface-induced moiré physics and to keep the continuum model analytically transparent.
The trilayer graphene Hamiltonian for the ABA and ABC stackings read
\begin{align}
    \mathbb{K}^{\mathrm{ABA}}_{3,\xi}=\begin{pmatrix}
        \mathbb{K}_{1,\xi}(\mathbf{k}) & \mathbb{U}_\xi(\mathbf{k}) & ~ \\
        \mathbb{U}^\dagger_\xi(\mathbf{k}) & \mathbb{K}_{1,\xi}(\mathbf{k}) & \mathbb{U}^\dagger_\xi(\mathbf{k}) \\
        ~ & \mathbb{U}_\xi(\mathbf{k}) &  \mathbb{K}_{1,\xi}(\mathbf{k})
    \end{pmatrix},\\\nonumber
    \mathbb{K}^{\mathrm{ABC}}_{3,\xi}=\begin{pmatrix}
        \mathbb{K}_{1,\xi}(\mathbf{k}) & \mathbb{U}_\xi(\mathbf{k}) & ~ \\
        \mathbb{U}^\dagger_\xi(\mathbf{k}) & \mathbb{K}_{1,\xi}(\mathbf{k}) & \mathbb{U}_\xi(\mathbf{k}) \\
        ~ & \mathbb{U}^\dagger_\xi(\mathbf{k}) &  \mathbb{K}_{1,\xi}(\mathbf{k})
    \end{pmatrix}.
\end{align}
The Hamiltonian of AB stacked bilayer graphene is
\begin{equation}
    \mathbb{K}^{\mathrm{AB}}_{2,\xi}=\begin{pmatrix}
        \mathbb{K}_{1,\xi}(\mathbf{k}) & \mathbb{U}_\xi(\mathbf{k})\\
        \mathbb{U}^\dagger_\xi(\mathbf{k}) & \mathbb{K}_{1,\xi}(\mathbf{k}) 
    \end{pmatrix}
\end{equation}
and $\mathbb{K}^{\mathrm{BA}}_{2,\xi}={\mathbb{K}^{\mathrm{AB}}_{2,\xi}}^\mathrm{T}$.
Here, $\mathbb{K}_1$ is the Hamiltonian of single layer graphene
\begin{equation}
    \mathbb{K}_{1,\xi}(\mathbf{k})=\begin{pmatrix}
        0 & \hbar v_0\Pi_\xi(\mathbf{k}) \\
        \hbar v_0\Pi^\dagger_\xi(\mathbf{k}) & 0
    \end{pmatrix},
\end{equation}
where $\Pi_\xi(\mathbf{k})=\xi k_x-ik_y$.
The interlayer hopping matrix is
\begin{equation}
    \mathbb{U}_{\xi}(\mathbf{k})=\begin{pmatrix}
        \hbar v_4\Pi_\xi(\mathbf{k}) & \hbar v_3\Pi^\dagger_\xi(\mathbf{k}) \\
        \gamma_1 & \hbar v_4\Pi_\xi(\mathbf{k})
    \end{pmatrix},
\end{equation} 
where $\hbar v_i=-\sqrt{3}\gamma_ia/2$ and $\gamma_i$ are the hopping parameters.
The tunneling matrix is given in real space by
\begin{equation}
\begin{aligned}
    \mathbb{T}_\xi(\mathbf{r})=
    &\begin{pmatrix}
    \gamma_{AA} &\gamma_{AB} \\
    \gamma_{AB} &\gamma_{AA}
    \end{pmatrix}+
    \begin{pmatrix}
        \gamma_{AA} &\gamma_{AB}\omega^{-\xi}\\
        \gamma_{AB}\omega^\xi & \gamma_{AA}
    \end{pmatrix}e^{-i\xi\mathbf{G_1}\cdot\mathbf{r}}
    +\\
    &\begin{pmatrix}
        \gamma_{AA} & \gamma_{AB}\omega^{\xi}\\
        \gamma_{AB}\omega^{-\xi} &\gamma_{AA}
    \end{pmatrix}e^{-i\xi(\mathbf{G_1+G_2})\cdot\mathbf{r}},
\end{aligned}
\end{equation}
where $\omega=e^{2\pi i/3}$, $\gamma_{AA}$ and $\gamma_{AB}$ are hopping amplitudes at $AA$ and $AB$ regions, respectively. We fit
$\{\gamma_0,\gamma_1,\gamma_2,\gamma_3,\gamma_4\}=\{-2.9, 0.39,0,0.2,0.15\}~\mathrm{eV}$, and $\gamma_{AA}=\gamma_{AB}=0.095~\mathrm{eV}$ by comparing with the TB results. The lattice relaxation is included by setting
$\gamma_{AA}=0.08~\mathrm{eV}$ and $\gamma_{AB}=0.11~\mathrm{eV}$. The trigonal warping terms $\gamma_3$ and $\gamma_4$ can significantly modify the spectrum near the charge neutrality point, which deserves explicit study \cite{zhou2023imaging}. In general, the band structure and topology of the twisted bilayer-trilayer are highly sensitive to the hopping parameters due to the moiré magnify effect \cite{su2025moire,phong2025coulombinteractionstabilizedisolatednarrow}. However, some features of the ABC cases are stable,  regardless of the hopping terms \cite{phong2025coulombinteractionstabilizedisolatednarrow}. We will discuss this point in details in the following section.

\subsection{Valley Chern Number}
We calculate the valley Chern number for each isolated band using the discretized Brillouin-zone method \cite{fukuiChernNumbersDiscretized2005}. 
In this approach, the Brillouin zone is discretized into a finite momentum-space mesh, and for the $n$th band the $U(1)$ link variable along the $\mu$ direction is defined as
\begin{equation}
U_{\mu}(\mathbf{k}) = \frac{\langle u_n(\mathbf{k}) \mid u_n(\mathbf{k}+\hat{\mu}) \rangle}{\left| \langle u_n(\mathbf{k}) \mid u_n(\mathbf{k}+\hat{\mu}) \rangle \right|},
\end{equation}
where $\ket{u_n(\mathbf{k})}$ is the cell-periodic Bloch eigenstate and $\hat{\mu}$ denotes a primitive step on the discretized momentum grid. 
The lattice Berry curvature on each plaquette is then given by
\begin{equation}
F_{12}(\mathbf{k}) = \ln \left[ U_1(\mathbf{k}) U_2(\mathbf{k}+\hat{1}) U_1^{-1}(\mathbf{k}+\hat{2}) U_2^{-1}(\mathbf{k}) \right],
\end{equation}
with the branch chosen such that $-\pi < \mathrm{Im}\,F_{12}(\mathbf{k}) \le \pi$. 
The valley Chern number is obtained by summing the lattice Berry curvature over the entire Brillouin zone
\begin{equation}
C_n = \frac{1}{2\pi i} \sum_{\mathbf{k}} F_{12}(\mathbf{k}).
\end{equation}

\subsection{Hartree Potential}\label{sec:Hartree_theory}

To include long-range Coulomb interactions at the mean-field level, we follow the self-consistent Hartree construction of Guinea \textit{et al.} \cite{guineaElectrostaticEffectsBand2018, ceaElectronicBandStructure2019}.
Neglecting higher-order harmonics (we have verified that higher order harmonics can be neglected), the Hartree potential takes the form  
\begin{equation}
v_H(\mathbf{r}) = V_0 \, \delta \rho_G \sum_i e^{i \mathbf{G}_i \cdot \mathbf{r}},
\label{eq:hartree}
\end{equation}
where
\begin{equation}
V_0 = \frac{e^2}{\epsilon |\mathbf{L}_i|},
\end{equation}
and the dominant contribution arises from the first shell of moiré reciprocal vectors
\begin{equation}
\mathbf{G}_i = \pm \mathbf{G}_1,\ \pm \mathbf{G}_2,\ \pm (\mathbf{G}_1 + \mathbf{G}_2).
\end{equation}
The single dimensionless parameter $\delta \rho_G$ is 
given by
\begin{equation}
    \delta \rho(\mathbf{G}) =
4 \int_{\mathrm{mBZ}} \frac{d^2 \mathbf k}{V_{\mathrm{mBZ}}}
\sum^{\prime}_{\mathbf{G}' i, n}
\phi^{i,*}_{n\mathbf{k}}(\mathbf{G}')
\phi^{i}_{n\mathbf{k}}(\mathbf{G}' + \mathbf{G}),
\end{equation}
where $V_{\mathrm{mBZ}}$ is the area of the mBZ, $i$ is the sublattice/layer index, $n$ is the moiré bands index, $\phi^{i}_{n\mathbf{k}}(\mathbf{G})$ is the amplitude for an electron to occupy a state with Bloch momentum $\mathbf{k}+\mathbf{G}$ in the $n$th band, and the primed summation implies taking only occupied (or unoccupied) states with respect to charge neutrality. Because of $C_3$ symmetry and the real $\delta\rho$, the components linked by $C_3$ rotation have the same values and the two groups should be conjugate.

The Hartree potential of Eq.~\eqref{eq:hartree} contributes to the CM Hamiltonian as a diagonal term in the sublattice/layer subspace, coupling the generic state of momentum $\mathbf{k}$ to the six states of momentum $\mathbf{k}-\mathbf{G}_i$, with an amplitude equal to $V_0\delta\rho_G$.
This yields a filling-dependent periodic electrostatic potential that couples states separated by moiré reciprocal lattice vectors and captures the leading Hartree-induced renormalization of the flat bands. In the numerical results, $\delta\rho(\mathbf{G})$ is determined iteratively until convergence, of which the criterion is defined as the error between two steps being smaller than $10^{-6}$. 
We consider two valleys and take $\epsilon=10$ as dielectric constant. 

We note that in the Hartree potential modeling, we only consider the electrostatic screening from the metallic gates. Recently, Phong \textit{et al.} \cite{phong2025coulombinteractionstabilizedisolatednarrow} used a layer-dependent Coulomb potential that considers both the screening effects from the dual gates and the charge redistribution among the multilayers to screen the applied displacement field, and found robust higher-Chern bands in twisted rhombohedral trilayer-bilayer graphene. We leave the investigation of the extended Hartree-Fock effect to the feature work.  

\begin{figure*}[!t]
    \centering
    \includegraphics[width=0.9\linewidth]{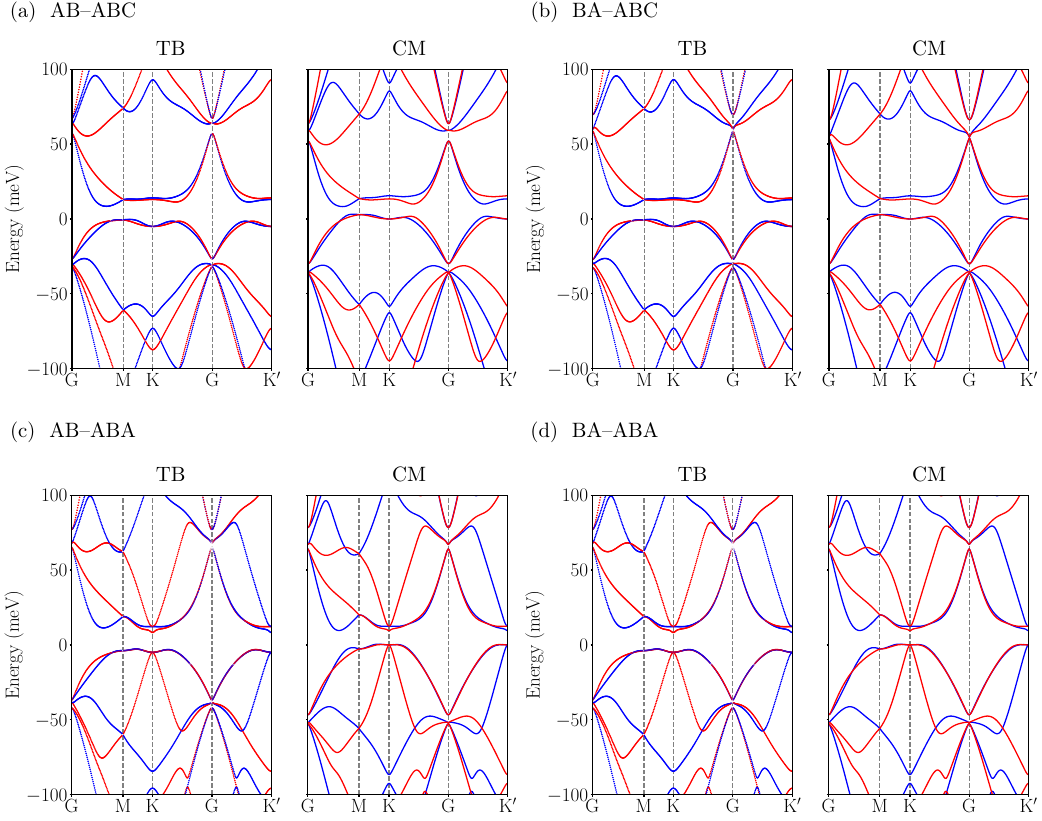}
    \caption{Low-energy band structures from the TB and CM approaches for the unrelaxed AB--ABC, BA--ABC, AB--ABA, and BA--ABA stackings at a twist angle of $1.54^\circ$. In the CM results, the $+1$ and $-1$ valleys are shown in red and blue, respectively. In the TB results, the valley character is resolved using the valley operator. The hopping terms in the CM model are adapted to have a good agreement of the band structures with the TB results. We set $\gamma_{AA}=\gamma_{AB}=95~\mathrm{meV}$.}
    \label{fig:band_compare}
\end{figure*}

\section{Results and discussion}
\label{sec:results}
\subsection{Band Structures}

\begin{figure*}
    \centering
    \includegraphics[width=\linewidth]{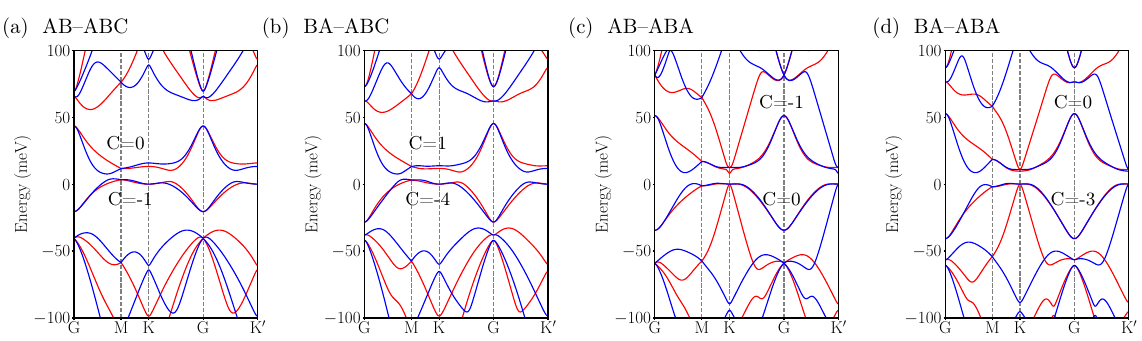}
    \caption{Low-energy CM band structures for relaxed AB--ABC, BA--ABC, AB--ABA, and BA--ABA stacking configurations at a twist angle of $\theta=1.54\degree$. The hopping terms in the CM approach are the same as in Fig. \ref{fig:band_compare}. We consider the lattice relaxation by resetting $\gamma_{AA}=80~\mathrm{meV}$ and $\gamma_{AB}=110~\mathrm{meV}$. The Chern numbers of the flat bands in the $+1$ valley are indicated.}
    \label{fig:relax_band}
\end{figure*}

We first investigate the band structure of twisted bilayer-trilayer graphene with four different arrangements via the TB and CM models. As shown in Fig.~\ref{fig:band_compare}, by manipulating the hopping terms in the CM, the two methods exhibit excellent agreement within the low-energy window in both valleys. 
Unlike in TBG, the band structures along $\mathrm{G} \rightarrow \mathrm{K}$ are nondegenerate, reflecting the absence of $C_{2x}$ symmetry. 
Despite the reduced symmetry, the moiré bands remain qualitatively similar to those of TBG, suggesting that the low-energy electronic structure is still primarily governed by the moiré coupling between the twisted bilayer subsystems.

A notable distinction arises for the AB-ABA and BA-ABA stackings, where an additional dispersive band touches the flat band at (K) in the (+1) valley and at (K´) in the (-1) valley. This band exhibits a gapped monolayer-Dirac-like dispersion and originates from the monolayer-like sector of the ABA trilayer \cite{minChiralDecompositionElectronic2008}, even if considering the relaxation effect as shown in Fig.~\ref{fig:relax_band}. 
In an isolated ABA trilayer, this sector is formed by an antisymmetric combination of the two outer layers and is decoupled from the bilayer-like sector by mirror symmetry. 
In the twisted five-layer structure, however, only one side of the ABA trilayer is coupled to the adjacent bilayer, so that the mirror symmetry is broken and the monolayer-like sector is no longer exactly decoupled. 
It consequently hybridizes with the moiré bands and becomes continuously connected to the moiré narrow bands. Its hybridization with the flat bands produces additional gap closing and reopening, and allows Chern number transfer between the flat and remote bands. In all the four stackings, a gap between the narrow bands appears in the non-interacting band structures, which is consistent with the theoretical work \cite{phong2025coulombinteractionstabilizedisolatednarrow} and the experimental results \cite{waters2024topological}.

The band structures of the parallel- and antiparallel-stacking configurations are otherwise qualitatively similar, and resemble the twisted double bilayer graphene cases \cite{koshinoBandStructureTopological2019}. 
Their differences arise primarily from the hybridization of the central conduction and valence bands with nearby remote bands, whereas the overall dispersion of the central moiré bands is only weakly modified. 
This similarity can be understood from the fact that the low-energy states of Bernal bilayer graphene reside predominantly on the non-dimer sites. Consequently, changing the stacking chirality interchanges the sublattices that form the vertically aligned dimer pair but has only a limited effect on the dispersion of the low-energy moiré bands.

By comparing the band structures in Figs. \ref{fig:band_compare} and \ref{fig:relax_band}, we see that the lattice relaxation flattens the narrow bands and opens a gap between the narrow and remote bands. Despite the relatively small changes in the conduction- and valence-band dispersions induced by an outer-layer translation, the corresponding Chern numbers can differ, as shown in Fig.~\ref{fig:relax_band}. The Chern number is determined by the Berry-curvature distribution over the entire mBZ. Moreover, for a fixed valley and a fixed convention for the layer-dependent potential, reversing the stacking chirality from AB to BA reverses the associated valley-topological contribution of the bilayer block. 
This reversal gives rise to the relative plus or minus sign appearing in the hierarchical Chern-number sum rules~\cite{liuQuantumValleyHall2019,ledwithFamilyIdealChern2022,wangHierarchyIdealFlatbands2022}. 
Consequently, two structures with nearly identical dispersions along high-symmetry lines may nevertheless carry different Chern numbers.

\begin{figure*}[t!]
    \centering
    \includegraphics[width=\linewidth]{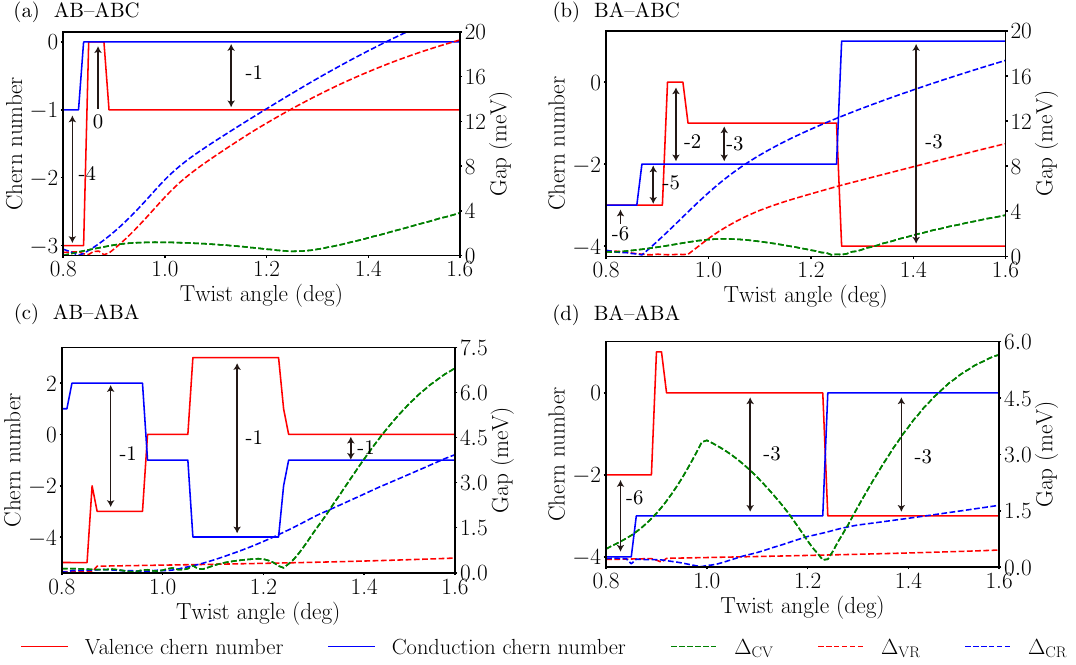}
    \caption{Chern numbers and energy gaps as functions of twist angle for four stacking configurations: (a) AB--ABC, (b) BA--ABC, (c) AB--ABA, and (d) BA--ABA. The red and blue solid lines, corresponding to the left axis, denote the Chern numbers of the valence and conduction flat bands, respectively. The dashed lines, corresponding to the right axis, show the relevant band gaps: the green dashed line denotes the conduction-valence narrow band gap $\Delta_{\mathrm{CV}}$, while the red and blue dashed lines denote the gaps between the valence and conduction flat bands and their nearest remote bands, labeled as $\Delta_{\mathrm{VR}}$ and $\Delta_{\mathrm{CR}}$, respectively. The black lines with arrows shows the sum of the Chern number of the conduction band and valence band. The CM hopping terms are the same as in Fig. \ref{fig:relax_band}. }
    \label{fig:chern_twist}
\end{figure*}

\subsection{Twist-tunable flat band and topology}

\begin{figure*}[!htbp]
    \centering
    \includegraphics[width=\linewidth]{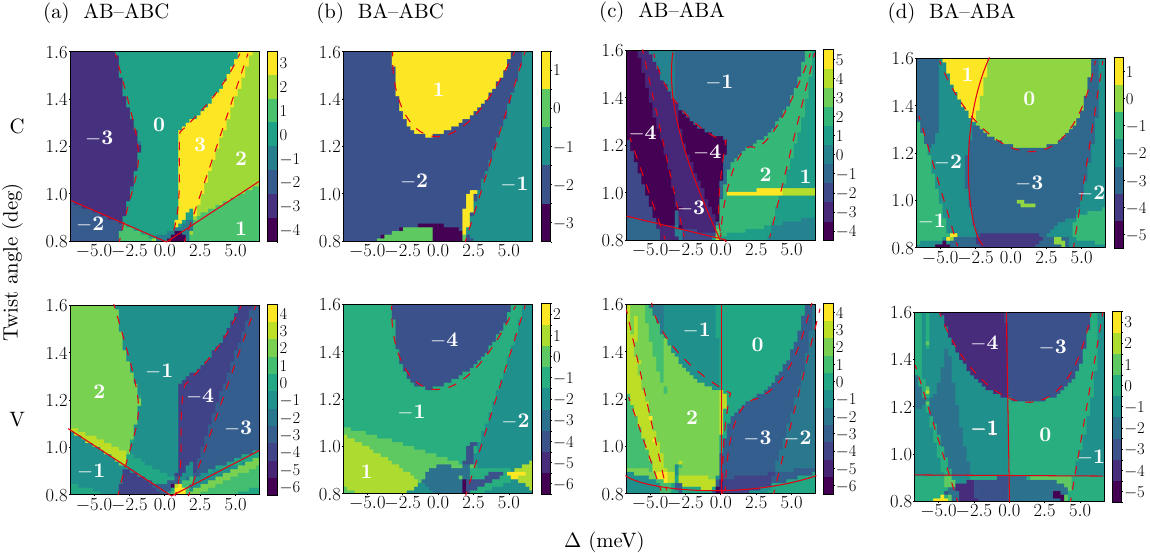}
    \caption{Chern-number phase diagrams of the conduction flat band (top row) and valence flat band (bottom row) as functions of twist angle $\theta$ and perpendicular electric field $E$ for (a) AB--ABC, (b) BA--ABC, (c) AB--ABA, and (d) BA--ABA stackings. The color scale represents the Chern number of the corresponding band. We label the Chern number for each color block. Red dashed lines indicate Type--I transition driven by the closing and reopening of the gap $\Delta_{\mathrm{CV}}$, whereas red solid lines indicate Type--II transition associated with the gap $\Delta_{\mathrm{CR}}$ or $\Delta_{\mathrm{VR}}$. Only the larger and well-isolated phase regions (can be easily recognized from the band gap) are labeled for clarity. A $100\times100$ $k$-point mesh is used in the Chern-number calculations. The CM hopping terms are the same as in Fig. \ref{fig:relax_band}.}
    \label{fig:chern_gate}
\end{figure*}

We next investigate the evolution of the Chern number as a function of twist angle via the CM approach. We plot both the Chern number and the relevant band gaps versus twist angle in Fig. \ref{fig:chern_twist}. This comparison allows us to directly identify which gap-closing and reopening processes are responsible for changes in the Chern number. 
In practice, in some cases, even with a dense $k$ mesh ($100 \times 100$), it can be difficult to determine whether an apparently vanishing gap is physically real or merely a numerical artifact, especially at small twist angles where narrow bands and remote bands compact in a narrow energy window. A robust topological phase requires both a finite gap between the conduction--valence narrow bands ($\Delta_{\mathbf{CV}}$) and a sufficiently large gap to the remote bands ($\Delta_{\mathbf{CR}}$/$\Delta_{\mathbf{VR}}$), i.e. an isolated band. Presenting the Chern number and the band-gap evolution in the same figure therefore provides a direct way to assess the type of topological transition.

Consistent with Ref.~\cite{liuQuantumValleyHall2019} and its interpretation in the chiral limit by Refs.~\cite{ledwithFamilyIdealChern2022,wangHierarchyIdealFlatbands2022}, we find that at relatively large twist angles ($\gtrsim 1^\circ$) the total Chern number of the two flat bands is $(M-N)$ for parallel stacking and $(M+N-2)$ for antiparallel stacking in all four arrangements. 
In the cases considered here, the sum of the Chern numbers of the conduction band (blue curves) and valence band (red curves) is $-1$ for the AB-bilayer-based structures and $-3$ for the BA-bilayer-based structures. 
While this overall trend remains unchanged, additional topological transitions occur in the BA-ABC, AB-ABA, and BA-ABA stackings around $1.25^\circ$. 
In these cases, the Chern numbers of the conduction and valence narrow bands change through gap-closing and gap-reopening processes of $\Delta_{\mathbf{CV}}$, as indicated by the green curves. 
At smaller twist angles, both the $\Delta_{\mathbf{CV}}$ and $\Delta_{\mathbf{CR}}$/$\Delta_{\mathbf{VR}}$ become very small, making it difficult to determine whether an apparent gap closing is physical or numerical. However, a change of Chern number would always imply a gap closing and reopening, and vice versa. 
The asymptotic Chern number rules derived in the chiral limit assume that the two flat bands remain isolated from the remote bands. 
Once the remote gap collapses, the two-band subspace is no longer well defined, and Chern number transfer between the flat and remote bands invalidates the simple counting rule. These cases therefore require careful examination, or alternatively the application of a vertical electric field to enlarge the relevant gaps and thereby clarify the topology. 

Therefore, with the evolution of the twist angle, there are two types of topological transition of the narrow bands: one appears between the narrow bands that follows the sum rules (Type--I); the other occurs between the narrow and adjacent remote bands that do not follow the sum rules (Type--II). Moreover, the Chern number of the narrow bands are more stable in relatively larger angles, which have large enough gap to isolate the narrow bands. In all the four stacking configurations, the Type--II transition takes place between a narrow band and its remote band, with the $\Delta_{\mathbf{VR}}/\Delta_{\mathbf{CR}}$ closing and reopening. The crossover angle is below 1$^\circ$, which highly depends on the strength of lattice relaxation \cite{guinea2019continuum}. Such topological transition was also observed in TBG with angle below 1$^\circ$~\cite{navarro2026topological}, and magic angle under heterostain  \cite{escudero2026straintronics}.

\begin{figure*}[!htbp]
    \centering
    \includegraphics[width=\linewidth]{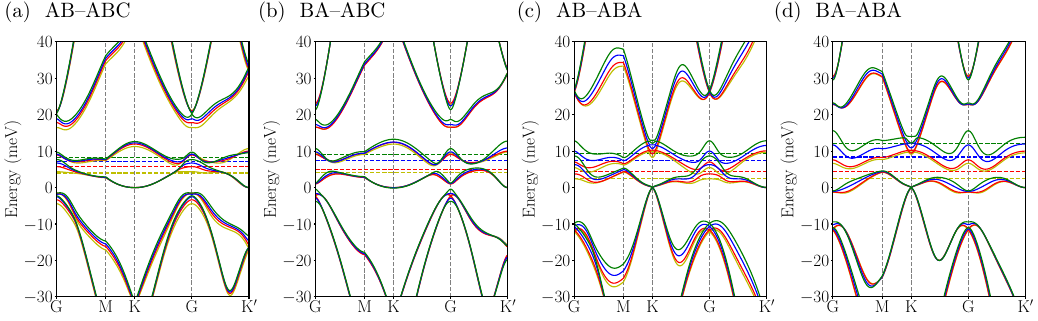}
    \caption{Hartree-corrected self-consistent band structures for four stacking configurations: (a) AB--ABC, (b) BA--ABC, (c) AB--ABA, and (d) BA--ABA at $\mathrm{twist}=1.05^\circ$ and zero electric field. The yellow, red, blue, and green bands correspond to fillings $\nu=-1$, $0$, $1$, and $2$, respectively. The horizontal dashed lines denote the corresponding Fermi energies.}
    \label{fig:hartree_band}
\end{figure*}

\subsection{Gate-tunable flat band and topology}

Next, we investigate the evolution of the Chern number under an external electric field applied perpendicular to the layers by introducing a layer-dependent onsite potential in the CM approach. The results can be seen in Fig.~\ref{fig:chern_gate}, which shows a two-dimensional phase diagram of the valley Chern number as a function of the electric field $\Delta$ and twist angle. The phase boundaries are line labeled according to the gap reopening mechanism responsible for the corresponding topological transition, i.e., red solid line for Type--I transition and red dashed line for Type--II transition.  We only mark the phase boundary in the relatively large twist angle region, of which the narrow bands have clear topological transitions. For angles lower than 0.9$^\circ$, the low energy bands are quite narrow with tiny gaps, which makes the change of the valley Chern number more complicated to resolve.

The conduction and valence band states are predominantly layer polarized~\cite{wangHierarchyIdealFlatbands2022}. 
Consequently, a layer-dependent potential can effectively isolate these bands or induce rehybridization. Again, two types of topological transitions are observed. In certain parameter regimes, these two transitions can occur simultaneously.
Note that because of the lack of $C_{2z}$ symmetry, the phase diagrams are not symmetric under electric field reversal $E\rightarrow-E$. Particularly, in the rhombohedral case with relatively large twist angle, the topology of the narrow bands are more stable under the negative displacement field, of which the rhombohedral trilayer subsystem plays a more dominant role. On the contrary, when the ABA stacking dominates, the Chern number of the narrow bands are more tunable under the negative electric field. In the four configurations under positive electric fields, the bilayer subsystem dominates the low energy band structure, resulting similar topological transitions.   

The four stacking configurations exhibit qualitatively distinct field-angle phase diagrams. In particular, the ABC-based structures generally display broader and cleaner topological regions, and have the Type--II transition at small twist angles. The antiparallel configuration hosts a large region of conduction narrow band with $C=-2$, and the parallel configuration with $C=-3$. Interestingly, these results are consistent with the Ref.~\cite{phong2025coulombinteractionstabilizedisolatednarrow}, which uses a different set of CM hopping terms. Recently, the conduction narrow band with $|C|=3$ in twisted rhombohedral-stacked trilayer-bilayer graphene with $\theta \approx1.5^\circ$ and under displacement field (field direction point to trilayer), has been experimentally observed \cite{dongObservationIntegerFractional2025}.
For the ABA-based structure, there is an extra nearly-vertical red solid line corresponding to Type--II transition, which causes a more fragmented pattern. 
This complexity is consistent with the additional monolayer-like band discussed above, which participates in field-tuned hybridization and generates extra gap closings and Chern number transfer processes. Recently, a large extent of symmetry-broken correlated states have been observed experimentally in the ABA stacking cases, showing the higher tunability of the system \cite{waters2024topological}. Moreover, the band structure and topology are highly sensitive to the CM hopping terms. For example, in the Ref.~\cite{waters2024topological} with different hopping terms, the CM non-interacting band structure has an overlap of the narrow bands at the charge neutrality point. Consequently, the experimentally observed insulating state at filling $\nu=0$ is argued to most likely be a correlated state. However, in our non-interacting band structures, we find a non-zero value of $\Delta_{\mathbf{CV}}$. We note that the sum rules are still valid for the Type--I transition under an electric field. The sum rules are derived from the Chiral limit, where the dominant parts are in the non-diagonal terms \cite{liuQuantumValleyHall2019,ledwithFamilyIdealChern2022,wangHierarchyIdealFlatbands2022}. The onsite potential terms contribute to the diagonal term in the CM Hamiltonian, which play no role in the sum rules.   

\begin{figure*}[t!]
    \centering
    \includegraphics[width=\linewidth]{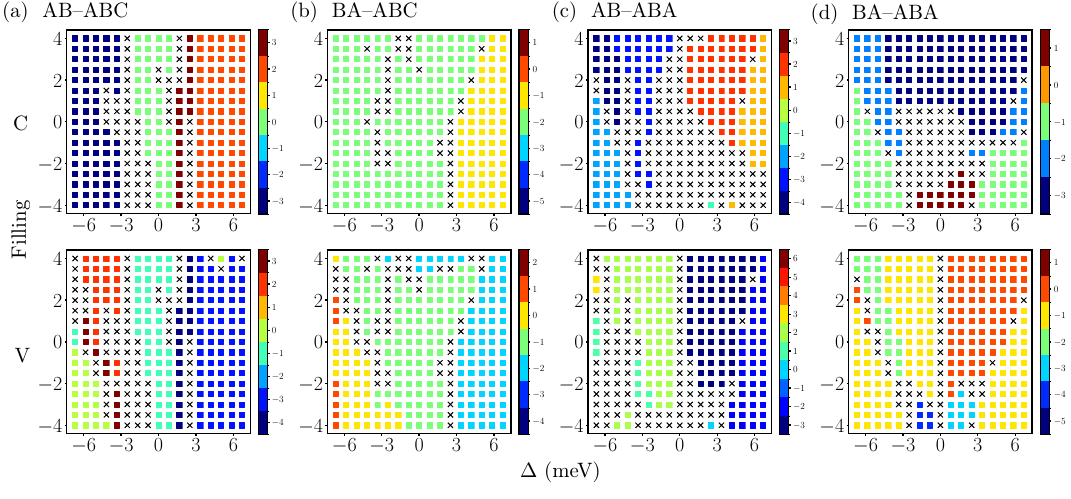}
    \caption{Chern-number maps of the conduction (top row) and valence (bottom row) flat bands as functions of the filling factor $\nu$ and perpendicular electric field $E$, for the (a) AB--ABC, (b) BA--ABC, (c) AB--ABA, and (d) BA--ABA stackings at a fixed twist angle of $\theta=1.05^\circ$. The color scale indicates the Chern number of the corresponding flat band. Colored squares represent the sampled points in the parameter space, while black crosses denote points at which the corresponding band is not isolated well from the other bands. We assume that the band is not isolated if the gaps with adjacent bands are smaller than 0.3~meV. The crosses in all configurations with $\nu=0$ are due to the value of gaps less than 0.3 meV.}
    \label{fig:hartree_map}
\end{figure*}

\subsection{Hartree-tunable flat band and topology}

We finally examine the effect of the self-consistent Hartree potential on the low-energy bands. For each filling, the Hartree potential is obtained iteratively within the first-harmonic approximation (enough to converge the calculations) described in Sec.~\ref{sec:Hartree_theory} and is then included as a moiré-periodic diagonal potential in the CM Hamiltonian. 

Figure~\ref{fig:hartree_band} shows the resulting self-consistent band structures for the four stacking configurations at fillings $\nu=-1$, $0$, $1$, and $2$ (corresponding to $\nu+4$ electrons per moiré unit cell). Compared with twisted bilayer graphene~\cite{guineaElectrostaticEffectsBand2018, ceaElectronicBandStructure2019}, the Hartree correction in twisted bilayer-trilayer systems is substantially weaker. 
The main effect is a filling-dependent shift (rigid shift) and a minor reshaping of the low-energy bands, while the overall miniband dispersion and bandwidth remain largely unchanged. 
This weak impact indicates that the first-harmonic component of the charge-density modulation is relatively small in these systems. 
Physically, this can be attributed to the fact that the low-energy wave functions are distributed over a larger number of layers and sublattices, thus reducing the moiré-scale charge inhomogeneity responsible for the Hartree potential.

The Hartree effect is more prominent in the ABA-based stackings than in the ABC-based stackings. 
This behavior is consistent with the additional monolayer-like band discussed above for the ABA trilayer.
Since this band lies close to the flat-band manifold, a weak filling-dependent electrostatic potential can modify its hybridization with the flat bands, leading to more pronounced changes in the band positions and avoided crossings. 
In contrast, the ABC-based stackings exhibit a cleaner isolated flat-band manifold and are therefore less sensitive to the self-consistent Hartree correction.

Despite these band-structure modifications, the Chern numbers of the conduction and valence flat bands remain unchanged for the fillings considered in Fig.~\ref{fig:hartree_band}. 
This is because the self-consistent Hartree potential does not close the relevant conduction--valence gap or the gap to the remote bands. 
The Hartree correction can redistribute the Berry curvature in momentum space, but without a gap-closing event the integrated Chern number is topologically protected. 

We further examine the combined effect of carrier filling and perpendicular electric field on the band topology at a fixed twist angle $\theta=1.05^\circ$. 
Figure~\ref{fig:hartree_map} shows the Chern-number maps of the conduction and valence flat bands in the $(\nu,\Delta)$ filling factor-electric field parameter space, for the four stacking configurations. 
Overall, the dominant variation of the Chern number is controlled by the electric field, while the dependence on filling is comparatively weak. 
This behavior is consistent with the Hartree-corrected band structures in Fig.~\ref{fig:hartree_band}, where the self-consistent Hartree potential mainly induces filling-dependent energy shifts and minor band reshaping, without substantially modifying the overall miniband dispersion.

The weak filling dependence indicates that the Hartree correction does not generally close the relevant gaps over most of the parameter space. 
As a result, the Chern numbers remain robust within extended regions of the $(\nu,\Delta)$ plane. Nevertheless, close to topological phase boundaries, the filling-dependent Hartree potential can shift the relative positions of the flat and remote bands and thereby move the gap-closing points. This gives rise to filling-induced changes of the Chern number in narrow regions of the phase diagram.

The stacking dependence is also evident. The ABC-based structures exhibit relatively regular and extended topological regions, whereas the ABA-based structures display a more fragmented pattern. 
This difference can be traced back to the additional monolayer-like band in the ABA trilayer, which lies close to the flat-band manifold and can be rehybridized by either the external electric field or the self-consistent Hartree potential.
Consequently, the ABA-based systems are more susceptible to additional gap closings and Chern-number transfer between the flat and remote bands.

\section{Conclusion}
\label{sec:conclusion}

In summary, we have systematically investigated the low-energy electronic structure and band topology of twisted bilayer--trilayer graphene with four stacking configurations: AB--ABC, BA--ABC, AB--ABA, and BA--ABA. 
By comparing atomistic TB calculations with CM results, we find good agreement in the band structures in the low-energy regime, demonstrating that the CM approach captures the essential moiré-band features of these five-layer systems. Although the four configurations exhibit qualitatively similar central narrow bands, their hybridization with nearby remote bands depends sensitively on the stacking sequence. 
In particular, the ABA-based structures host an additional monolayer-like band originating from the ABA trilayer, which enhances hybridization with the central bands and gives rise to additional gap closings and topological transitions. 
We find that the hierarchical Chern number sum rules are generally satisfied only at relatively large twist angles (above magic angle) and under external electric field, where the Type--I transitions occur. We also find a different Type--II transition between the narrow band and its adjacent remote band.
We further investigate the evolution of the band topology as functions of twist angle, perpendicular electric field, and carrier filling within the self-consistent Hartree approximation. 
The obtained results reflect that the twist angle and perpendicular electric field provide effective means of tuning the band topology, owing to the strong layer polarization of the low-energy states. The Hartree potential mainly induces moderate energy shifts and weak reshaping of the narrow bands, without triggering topological transitions. However, at non-zero displacement field, the narrow bands and its topology are well tuned by the Hartree potential. Our results demonstrate that the ABA and ABC stacking based twisted bilayer-trilayer graphene have different responses to the tuning knobs, making this moiré structure a more versatile system to generate correlated and topological features.\\

\begin{acknowledgments}
F.E. and Z.Z. acknowledge support from NOVMOMAT, Grant PID2022-142162NB-I00 funded by MCIN/AEI/ 10.13039/501100011033 and, by “ERDF A way of making Europe”, and from the ‘Severo Ochoa’ Programme for Centres of Excellence in R\&D (CEX2020-001039-S/AEI/10.13039/501100011033). F.E. acknowledges support funding from the European Union's Horizon 2020 research and innovation programme under the Marie Skłodowska-Curie grant agreement No 101210351. Z.Z acknowledges support from the European Union's Horizon 2020 research and innovation programme under the Marie-Sklodowska Curie grant agreement No 101034431, which funded partially the reserach presented in this work. S.Y. acknowledges funding from the National Natural Science Foundation of China (Grants No. 12425407). 
Numerical calculations presented in this paper have been performed in the Supercomputing Center of Wuhan University. 
\end{acknowledgments}

\bibliography{newref.bib}

\end{document}